\documentstyle[12pt]{article}
\newcommand{\be}[1]{\begin{equation} \label{(#1)}}
\newcommand{\ee}{\end{equation}}
\newcommand{\ba}[1]{\begin{eqnarray} \label{(#1)}}
\newcommand{\ea}{\end{eqnarray}}
\newcommand{\nn}{\nonumber}
\newcommand{\rf}[1]{~(\ref{(#1)})}
\begin{document}
\bigskip
\begin{center}
{\bf Possible constraints\\ on SUSY-model Parameters
from\\ Direct Dark Matter Search
\footnotemark[1]. }

\bigskip
{V.A. Bednyakov, H.V. Klapdor-Kleingrothaus$^*$, S.G. Kovalenko
\bigskip

{\it Joint Institute for Nuclear Research, Dubna, Russia}
\bigskip

$^*${\it
Max-Plank-Institut f\"{u}r Kernphysik, D-6900, Heidelberg,
Germany}
}
\end{center}
\begin{abstract}
        We consider the SUSY-model neutralino as a dominant Dark Matter
        particle in the galactic halo and investigate some general issues
        of direct DM searches via elastic neutralino-nucleus scattering.
        On the basis of conventional assumptions about the nuclear and
        nucleon structure, without referring to a specific SUSY-model,
        we prove that it is impossible in principle to extract more
        than three constraints on fundamental SUSY-model parameters
        from the direct Dark matter searches.
        Three types of Dark matter detectors probing different groups of
        parameters are recognized.
\end{abstract}
\bigskip
\footnotetext[1]{
This work was supported in part by the Russian Foundation for Fundamental
Research (93-02-3744).
}
\newpage
        Many physicists at present believe that  more than $90\%$ of the
        total universe mass should be in the form of non-luminous
        non-baryonic dark matter (DM).
        Most of our galaxy mass should be in a dark halo as well.
        According to the commonly accepted halo model the DM particles
        have a Maxwellian velocity distribution in the galactic frame,
        a mass density in the Solar system of about
        $\rho\approx$ 0.3 GeV$\cdot$cm$^{-3}$ and come to the Earth with mean
        velocities $v\approx$ 320 km/sec, producing a substantial flux
        $\Phi = \rho\cdot v/M$ ($\Phi>10^{7}$ cm$^{-2}$ sec$^{-1}$ for
        a particle mass $M\sim$ 1 GeV).
        Thus the idea of detecting DM particles directly,
        for instance through the elastic scattering off nuclei in detector
        target material looks realistic.

        Dark Matter in the universe consists of two main components,
        which play different cosmological roles.
        These are "hot" (HDM) and "cold" (CDM) DM components which decoupled
        from the thermal equilibrium in the early universe at relativistic
        and non-relativistic temperatures, respectively.
        The analysis of the COBE satellite data \cite{COBE} on anisotropy
        in the cosmic background radiation and the theory of
        the formation of large scale structure in the universe lead to the
        following DM composition
        DM = 0.7(CDM) + 0.3(HDM) \cite{Taylor}, \cite{Davis}.
        It is commonly believed that the CDM is represented by weakly
        interacting massive particles (WIMP) and the axion,
        while the HDM consists of light neutrinos.
        We do not discuss the axion as a DM candidate,
        concentrating on the pure WIMP variant of the CDM.

        The neutralino ($\chi$) is a SUSY-model version of WIMP.
        Presently it is the most favorable CDM candidate.
        $\chi$ is a Majorana ($\chi^{c}=\chi$) particle with
        spin~$\frac{1}{2}$ and usually assumed to be the lightest
        SUSY-particle (LSP).

        The problem of direct detection of the DM neutralino $\chi$
        via elastic scattering off nuclei has been considered by
        many authors and remains a field of great experimental
        and theoretical activity \cite{Witt}-\cite{BKK}.

        One can meet in the literature the statement that the neutralino
        prefers to scatter off spin-non-zero (J$\neq$0) nuclei due to
        its Majorana nature.
        However, we have shown \cite{BKK} that for sufficiently heavy nuclei
        nuclear spin is a non-dominant factor in the neutralino-nucleus
        scattering.
        Thus, one has equal chances to detect the DM signal using either
        spin-zero or spin-non-zero target nuclei with atomic weight $A>50$.

        However, J$\neq$0 and J=0 - nuclei provide different information
        about SUSY-model parameters.
        Therefore experiments with J$\neq$0 and with J=0 - nuclei are
        both important from the point of view of SUSY-phenomenology.

        In the present paper we envisage the question of how many independent
        constraints on fundamental SUSY-model parameters can be extracted
        from direct DM neutralino searches.
        On the basis of conventional assumptions we shall prove that there
        are just three independent constraints for special groups of
        parameters.

        In this respect we can distinguish three types of DM detectors
        with different sensitivity to the groups of fundamental parameters.
        These are detectors with  non-zero-spin nuclei of
        neutron-like (${}^{3}$He, ${}^{29}$Si, ${}^{73}$Ge,...) and
        proton-like (${}^{19}$F, ${}^{35}$Cl, ${}^{205}$Tl,...)
        shell-model structure, and spin-zero nuclei (${}^{76}$Ge,...).

        The DM event is elastic neutralino-nucleus scattering causing
        a nuclear recoil detected by detector.
        The event rate per unit mass of target material depends on the
        density and velocity of DM in the solar vicinity and the cross
        section $\sigma_{el}(\chi A)$ of the neutralino-nucleus
        ($\chi A$) elastic scattering.
        One can calculate $\sigma_{el}(\chi A)$ starting from the
        neutralino-quark effective Lagrangian.
        It can be written in the general form as
\be{Lagr} 
 L_{eff} = \sum_{q}^{}{\cal A}_{q}\cdot\bar\chi\gamma_\mu
            \gamma_5\chi\cdot
            \bar q\gamma^\mu\gamma_5 q +
    \frac{m_q}{M_{W}} \cdot{\cal S}_{q}\cdot\bar\chi\chi\cdot\bar q q,
\ee
        where terms with the vector and pseudoscalar quark currents are
        omitted being negligible in the case of non-relativistic
        DM neutralinos with typical velocities $v_\chi\approx 10^{-3} c$.

        The coefficients ${\cal A}_{q}, {\cal S}_{q}$ are effective
        phenomenological parameters completely parametrizing low-energy
        neutralino-quark interactions.
        They can be calculated in a specific SUSY-model.
        Furtheron we will consider the most general properties of the
        $\chi$-$A$ scattering starting from the effective Lagrangian\rf{Lagr}
        without referring to specific SUSY-models.
        To calculate the elastic neutralino-nucleus cross section
        $\sigma_{el}(\chi A)$ we follow the standard procedure of
        sequentially averaging the $\chi$-$q$ interactions over the nucleon
        and then the nuclear structure.

        The first and the second terms in $L_{eff}$\rf{Lagr} averaged
        over the nucleon states give spin-dependent and spin-independent
        matrix elements $M_{sd}$ and $M_{si}$, respectively.
        For the spin-dependent matrix element we have \cite{Witt},\cite{EF2}:
\be{Msd} 
    M_{sd}^{p(n)} = 4 \vec{S_\chi} \vec{S}_{p(n)}\cdot{\cal W}_{p(n)},
\ee
        where $\vec{S_{\chi}} \mbox{ and } \vec{S}_{p(n)}$ are
        the neutralino and proton (neutron) spin-operators.
        The factor ${\cal W}_{p(n)}$ depending on the parameters
        of the effective Lagrangian\rf{Lagr}  has the form
\be{Wsd}  
   {\cal W}_{p(n)} = \sum_{q \in p(n)}^{} {\cal A}_{q} \Delta q,
\ee
        $\Delta q$ are the fractions of the nucleon spin carried by the
        quark $q$.
        The definition is
\be{spin} 
   <p(n)|\bar q\gamma^\mu\gamma_5 q|p(n)> = 2 S_{p(n)}^{\mu} \Delta q,
\ee
    where $S_{p(n)}^{\mu}=(0,\vec{S}_{p(n)})$ is the 4-spin of the nucleon.
        The parameters $\Delta q$ for the proton can be extracted from
        the EMC \cite{EMC} and hyperon data \cite{Hyp}:
\be{spin_fract} 
        \Delta u =  0.77 \pm 0.08, \ \
        \Delta d = -0.49 \pm 0.08, \ \
        \Delta s = -0.15 \pm 0.08.
\ee
        The relevant values for the neutron can be found from \rf{spin_fract}
        by isospin symmetry substitution $u\rightarrow d, d\rightarrow u$.

        The spin-independent matrix element $M_{si}$ has the
        form \cite{Gelm},\cite{FOT}\footnote{ When this paper had been
          completed we received
        a paper ref.\cite{Drees} with more refined treatment of
        the spin-independent matrix element.}:
\ba{Msi} 
       M_{si} &=& \frac{M_{p(n)}}{M_{W}}\cdot\bar \chi\chi\cdot
       \bar \Psi \Psi\cdot {\cal V}.
\ea
        Again as in the spin-dependent matrix element\rf{Msd}
        we separate the factor
\be{Vsi} 
        {\cal V} = \sum_{q}^{} f_{q}\cdot {\cal S}_{q},
\ee
        where the parameters $f_q$ are defined as follows:
\be{Scal} 
   <p(n)|m_{q}\cdot \bar{q} q|p(n)> = f_{q}\cdot M_{p(n)}\bar \Psi \Psi.
\ee
       The values extracted from the data are \cite{Cheng},\cite{Gasser}:
$$
f_{u} = 0.022, \ \ \ \ \ f_{d} = 0.034, \ \ \ \ \ f_{s} = 0.38,
$$
$$
f_{c} = f_{b} = f_{t} = \frac{2}{27}(1 - f_{u} - f_{d} - f_{s}).
$$
        The next step is the calculation of nuclear matrix elements of scalar
        and axial-vector nucleon operators
        $\bar \Psi\cdot\Psi$ and $\bar \Psi\gamma_{\mu}\gamma_{5}\Psi$,
        respectively for the nuclear state $|A>$.
        The axial-vector operator corresponding in a non-relativistic
        limit to the nucleon spin-operator appears in eqn.\rf{Msd}.
        We can parametrize these nuclear matrix elements in the most
        general form as follows
\ba{Nucl} 
   <A|M_{p(n)}\bar \Psi \Psi|A> &=& M_{A} F_{si}(q^2) \bar{A}A,  \\
\nn
   <A|\vec{S}_{p(n)}|A> &=& \lambda_{p(n)} F_{sd}^{p(n)}(q^2) <A|\vec{J}|A>.
\ea
        Here $\vec{J}$ is the nuclear spin.
        The  form factors $F_{sd,si}(q^2)$ take into account effects of
        finite momentum transfer $q$ and obey the normalization conditions:
\be{norm} 
        F_{sd,si}(0) = 1.
\ee
        For nuclei with atomic weight $A>50$ nuclear structure is essential
        since $q_{max} \geq 1/R_{A}$
        ($R_{A}$ is the nuclear radius) \cite{EV}.
        In this case the form factors strongly deviate from the
        normalization value.

        The phenomenological parameter $\lambda_{p(n)}$ measures the
        contribution of the proton (neutron) group to the nuclear
        spin (See \cite{EV} for these nomenclature.).
        According to the conventional shell model picture, the odd-group
        contribution dominates for even-odd nuclei.
        For odd-proton group nuclei (${}^{7}$Li, ${}^{19}$F,
        ${}^{203}$Tl,...)  $\lambda_{p} >> \lambda_{n}$
        and for odd-neutron group nuclei (${}^{3}$He, ${}^{29}$Si,
        ${}^{73}$Ge,...) $\lambda_{n} >> \lambda_{p}$.

        In the case of odd-odd nuclei (${}^{10}$B, ${}^{14}$N,...)
        proton and neutron groups give comparable contributions to a
        nuclear spin and $\lambda_{n} \sim \lambda_{p}$.

        The $\lambda$-parameters should be calculated in some nuclear model.
        In the semi-empirical odd-group shell model \cite{EV} they can be
        related to the nuclear magnetic moment, $\mu$, as follows
\ba{lamb} 
   \lambda J = \frac{\mu - g^{l}J}{g^{s} - g^{l}},
\ea
        where $g^{l} = 1(0)$ and $g^{s} = 5.586(-3.826)$ are orbital and spin
        proton (neutron) $g$-factors.

        Then  one can extract $\lambda$-values from the experimental data
        on nuclear magnetic
        moments\footnote{Another approach based on the theory
        of finite Fermi systems is described in \cite{Nikolaev}.}.

        For our purpose the numerical values of these parameters are
        irrelevant but for completeness we give them in the Appendix
        for nuclei of interest in the DM search.

        Making standard transformations of the above-written matrix
        elements and averaging over the Maxwellian distribution of
        neutralino velocities, we finally arrive at the formula for the
        event rate of the elastic neutralino-nucleus scattering in the
        detector per day per unit mass of the target material:
\be{Rate} 
   R = \Bigl[ a_{p}\cdot{\cal W}_{p}^2 + a_{n}\cdot{\cal W}_{n}^2
          + a_{0}\cdot {\cal V}^2 \Bigr]
         \frac{\mbox{events}}{\mbox{kg}\cdot\mbox{day}}.
\ee
        The parameters $a_{i}$ depend on properties of the target nucleus
        as well as on the mass density and the average velocity of
        DM particles in the solar vicinity.
        In the Appendix we give the values of these parameters for
        different nuclei within  specific model assumptions.
        The above-defined quantities ${\cal W}$ and ${\cal V}$ contain all
        dependence on the parameters of the effective Lagrangian\rf{Lagr}
        and do not depend on nuclear properties.
        This factorization is a central point of the present paper.

        It follows from eqn.\rf{Rate} that measuring the event rate $R$
        we can study just three special combinations of fundamental
        parameters ${\cal W}_{p,n}$ and ${\cal V}$ defined in
        eqn.\rf{Wsd},\rf{Vsi}.
        This is the only information about  fundamental parameters accessible
        in DM search experiments.
        R is a linear combination of the quantities ${{\cal W}^2}_{p,n}$
        and ${\cal V}^2$.
        To extract experimental limitations for each of them one should
        search for DM with different target nuclei.
        We can distinguish three types of DM detectors
        by their sensitivity to ${\cal W}_{p,n}$ and ${\cal V}$.
        These are detectors built of:
\begin{enumerate}
\item spin-non-zero target nuclei with an 'odd-proton group'
        probing a linear combination
$$
 a_{p}\cdot{\cal W}_{p}^2 + a_{0}\cdot {\cal V}^2 < R^{(1)}_{exp}
$$
        and giving an experimental limit $ R^{(1)}_{exp}$;
\item spin-non-zero target-nuclei with an 'odd-neutron group' probing
        another linear combination
$$
   a_{n}\cdot{\cal W}_{n}^2 + a_{0}\cdot {\cal V}^2 <  R^{(2)}_{exp}
$$
        and giving an experimental limit $ R^{(2)}_{exp}$;
\item  spin-zero target nuclei sensitive only to the scalar part of
        the neutralino-nucleus interaction
$$
{\cal V}^2 <  R^{(3)}_{exp}
$$
        giving an experimental limit $ R^{(3)}_{exp}$.
\end{enumerate}

        In conclusion we would like to stress the following.
        To extract all information about SUSY-model parameters possible
        from direct DM search one should have three established types
        of DM detectors.
        No other information can be obtained from the direct
        DM search experiments.
        Different detectors can only improve the data on the three
        above-defined groups of SUSY-model parameters.

        A detailed discussion on the extractable information about
        SUSY model parameters will be given elsewhere.

\begin{table}[t]
\centerline{Table: Parameters for event rate calculations.}
\vspace*{5mm}
\begin{tabular}{|r|c|l|l|c||r|c|l|l|c|}\hline
{\small Isotope}&~~J~~&  $\lambda^2_p$ & $\lambda^2_n$ &
$\frac{r_{spin}}{r_{charge}}$ &
{\small Isotope}&~~J~~&  $\lambda^2_p$ & $\lambda^2_n$ &
$\frac{r_{spin}}{r_{charge}}$ \\ \hline \hline
$^{  1}$H  &  1/2 &  1.0    & 0      & 1.00 &$^{ 99}$Ru &  5/2 &
  0      & 0.0045 & 1.19 \\ \hline
$^{  3}$He &  1/2 &  0      & 1.2373 & 1.00 &$^{101}$Ru &  5/2 &
  0      & 0.0056 & 1.19 \\ \hline
$^{  7}$Li &  3/2 &  0.1096 & 0      & 1.17 &$^{107}$Ag &  1/2 &
  0.0720 & 0      & 1.06 \\ \hline
$^{  9}$Be &  3/2 &  0      & 0.0768 & 1.12 &$^{109}$Ag &  1/2 &
  0.0760 & 0      & 1.06 \\ \hline
$^{ 11}$B  &  3/2 &  0.0299 & 0      & 1.09 &$^{111}$Cd &  1/2 &
  0      & 0.0960 & 1.17 \\ \hline
$^{ 15}$N  &  1/2 &  0.1160 & 0      & 1.06 &$^{113}$Cd &  1/2 &
  0      & 0.1053 & 1.17 \\ \hline
$^{ 17}$O  &  5/2 &  0      & 0.0391 & 1.25 &$^{115}$Sn &  1/2 &
  0      & 0.2307 & 1.16 \\ \hline
$^{ 19}$F  &  1/2 &  0.8627 & 0      & 1.21 &$^{117}$Sn &  1/2 &
  0      & 0.2733 & 1.16 \\ \hline
$^{ 23}$Na &  3/2 &  0.0109 & 0      & 1.16 &$^{121}$Sb &  5/2 &
  0.0057 & 0      & 1.15 \\ \hline
$^{ 27}$Al &  5/2 &  0.0099 & 0      & 1.13 &$^{123}$Sb &  7/2 &
  0.0035 & 0      & 1.15 \\ \hline
$^{ 29}$Si &  1/2 &  0      & 0.0840 & 1.12 &$^{127}$I  &  5/2 &
  0.0026 & 0      & 1.15 \\ \hline
$^{ 31}$P  &  1/2 &  0.0760 & 0      & 1.11 &$^{129}$Xe &  1/2 &
  0      & 0.1653 & 1.14 \\ \hline
$^{ 35}$Cl &  3/2 &  0.0096 & 0      & 1.10 &$^{131}$Xe &  3/2 &
  0      & 0.0147 & 1.14 \\ \hline
$^{ 47}$Ti &  5/2 &  0      & 0.0067 & 1.20 &$^{133}$Cs &  7/2 &
  0.0033 & 0      & 1.14 \\ \hline
$^{ 49}$Ti &  7/2 &  0      & 0.0068 & 1.20 &$^{139}$La &  7/2 &
  0.0020 & 0      & 1.13 \\ \hline
$^{ 51}$V  &  7/2 &  0.0106 & 0      & 1.19 &$^{155}$Gd &  3/2 &
  0      & 0.0021 & 1.22 \\ \hline
$^{ 55}$Mn &  5/2 &  0.0069 & 0      & 1.17 &$^{157}$Gd &  3/2 &
  0      & 0.0035 & 1.22 \\ \hline
$^{ 59}$Co &  7/2 &  0.0049 & 0      & 1.15 &$^{183}$W  &  1/2 &
  0      & 0.0040 & 1.19 \\ \hline
$^{ 67}$Zn &  5/2 &  0      & 0.0083 & 1.13 &$^{191}$Ir &  3/2 &
  0.0387 & 0      & 1.08 \\ \hline
$^{ 69}$Ga &  3/2 &  0.0056 & 0      & 1.13 &$^{193}$Ir &  3/2 &
  0.0379 & 0      & 1.08 \\ \hline
$^{ 71}$Ga &  3/2 &  0.0237 & 0      & 1.13 &$^{199}$Hg &  1/2 &
  0      & 0.0693 & 1.17 \\ \hline
$^{ 73}$Ge &  9/2 &  0      & 0.0026 & 1.25 &$^{201}$Hg &  3/2 &
  0      & 0.0096 & 1.17 \\ \hline
$^{ 79}$Br &  3/2 &  0.0077 & 0      & 1.11 &$^{203}$Tl &  1/2 &
  0.2400 & 0      & 1.07 \\ \hline
$^{ 81}$Br &  3/2 &  0.0125 & 0      & 1.11 &$^{205}$Tl &  1/2 &
  0.2467 & 0      & 1.07 \\ \hline
$^{ 91}$Zr &  5/2 &  0      & 0.0186 & 1.21 &$^{207}$Pb &  1/2 &
  0      & 0.0960 & 1.17 \\ \hline
$^{ 93}$Nb &  9/2 &  0.0065 & 0      & 1.20 &$^{209}$Bi &  9/2 &
  0.0002 & 0      & 1.16 \\ \hline
\end{tabular}
\end{table}

\section{Appendix}
        Below we give complete formulae for calculation
        of the nuclear structure parameters entering in the definition of
        the event rate $R$  in eqn.\rf{Rate} within commonly accepted
        model assumptions.

        Assuming a gaussian parametrization \cite{Gauss} for the form
        factors $F_{sd,si}(q^2)$ in eqns.\rf{Nucl} and averaging over
        the Maxwellian distribution of DM particle velocities we get
\ba{abc} 
a_{p(n)} &=& 5.8\cdot 10^{10}\cdot\lambda^2_{p(n)}J(J + 1)\cdot
             \zeta(r_{spin})\cdot{\cal D},\\   
    a_0 &=& 1.44\cdot 10^{10}\cdot \bigl(\frac{M_A}{M_W}\bigr)^2
        \zeta(r_{charge})\cdot{\cal D},
\ea
        where the DM factor is
\ba{Dark} 
    {\cal D} &=& \Bigl[ \frac{4M_\chi M_A}{(M_\chi + M_A)^2}  \Bigr]
                 \Bigl[ \frac{\rho}{ .3 GeV\cdot cm^{-3} }    \Bigr]
                 \Bigl[ \frac{<|\vec{v_E}|>}{ 320 km/s }      \Bigr].
\ea
        Here $M_{\chi}, M_A$ are the DM neutralino and target nucleus masses.
        The coherence loss factor \cite{Gould} is defined by
\be{Corr}
        \zeta(r) = \frac{0.573}{b}
               \biggl(
            1 - \frac{\exp\bigl(-\frac{b}{1+b} \bigr)}{\sqrt{1+b}}
  \frac{ \mbox{erf} \bigl( \sqrt{ \frac{1}{1+b} } \bigr) }{\mbox{erf}(1)}
               \biggr),
\ee
        where
$$
        b = \frac{8}{9} \sigma^2 r^2
        \frac{ M_\chi^2 M^2_{A} }{ (M_\chi + M_{A})^2 }.
$$
        Here $\sigma^2$ is  the dispersion of the Maxwellian neutralino
        velocity distribution $\sigma = 0.9 \cdot 10^{-3}$.
        $r_{spin}$ and $r_{charge}$ are the $rms$ spin and charge
        radius of the nucleus $A$.
        To estimate $r_{charge}$ one can use
        the parametrization \cite{Ede}:
\be{rms}   
        r_{charge} = (0.3 + 0.89 M_{A}^{1/3}) \mbox{ fm}.
\ee
        The values of the ratio  $r_{spin}/r_{charge}$ can be estimated
        in the harmonic oscillator shell model \cite{EF1}.

        Numerical values of the above defined parameters for nuclei
        of interest in direct DM searches are given in the Table.

\end{document}